\documentclass{pasj00}

\def\Umlaut#1{\"{#1}}

\def\commenta{$^*$}
\def\commentb{$^\dagger$}
\def\commentc{$^\ddagger$}
\def\commentd{$^\S$}

\def\inpress{in press}
\def\astroph#1{ (astro-ph/#1)}

\DeclareAbbreviation\AAHam{Astron. Abh. Hamburg. Sternw.}
\DeclareAbbreviation\AARv{Astron. Astrophys. Rev.}
\DeclareAbbreviation\an{Astron. Nachr.}
\DeclareAbbreviation\AcA{Acta Astron.}
\DeclareAbbreviation\Afz{Astrofizika}
\DeclareAbbreviation\AnTok{Tokyo Astron. Obs. Annals, Sec. Ser.}
\DeclareAbbreviation\Ap{Astrophysics}
\DeclareAbbreviation\ARep{Astron. Rep.}
\DeclareAbbreviation\ATel{Astronomer's Telegram}
\DeclareAbbreviation\ATsir{Astron. Tsirk.}
\DeclareAbbreviation\AcApS{Acta Astrophys. Sinica}
\DeclareAbbreviation\AstL{Astron. Letters}
\DeclareAbbreviation\BaltA{Baltic Astron.}
\DeclareAbbreviation\BASI{Bull. Astron. Soc. India}
\DeclareAbbreviation\BeSN{Be Star Newsletter}
\DeclareAbbreviation\GCN{GCN}
\DeclareAbbreviation\ibvs{Inf. Bull. Variable Stars}
\DeclareAbbreviation\JAD{J. Astron. Data}
\DeclareAbbreviation\JAVSO{J. American Assoc. Variable Star Obs.}
\DeclareAbbreviation\JBAA{J. British Astron. Assoc.}
\DeclareAbbreviation\LowOB{Lowell Obs. Bull.}
\DeclareAbbreviation\MitVS{Mitteil. Ver\"{a}nderl. Sterne}
\DeclareAbbreviation\MmSAI{Mem. Soc. Astron. Ita.}
\DeclareAbbreviation\Msngr{Messenger}
\DeclareAbbreviation\NewA{New Astron.}
\DeclareAbbreviation\NewAR{New Astron. Rev.}
\DeclareAbbreviation\OAP{Odessa Astron. Publ.}
\DeclareAbbreviation\Obs{Observatory}
\DeclareAbbreviation\PASA{Publ. Astron. Soc. Australia}
\DeclareAbbreviation\PAZh{Pis'ma AZh}
\DeclareAbbreviation\PhR{Phys. Rep.}
\DeclareAbbreviation\PVSS{Publ. Variable Stars Sect. R. Astron. Soc. New Zealand}
\DeclareAbbreviation\PZ{Perem. Zvezdy}
\DeclareAbbreviation\PZP{Perem. Zvezdy Pril.}
\DeclareAbbreviation\QJRAS{QJRAS}
\DeclareAbbreviation\RMxAA{Rev. Mexicana Astron. Astrof.}
\DeclareAbbreviation\RvMA{Reviews of Modern Astron.}
\DeclareAbbreviation\Sci{Science}
\DeclareAbbreviation\SvA{Soviet Astronomy}
\DeclareAbbreviation\SvAL{Soviet Astronomy Letters}
\DeclareAbbreviation\VeSon{Ver\"{o}ff. Sternw. Sonneberg}
\DeclareAbbreviation\VSOLJBul{VSOLJ Variable Star Bull.}
\DeclareAbbreviation\yCat{VizieR Online Data Catalog}
\DeclareAbbreviation\ZA{Z. Astrophys.}

\def\ASPConf#1#2{ASP Conf. Ser. #1, #2}
\def\IAUColloq#1#2{IAU Colloq. #1, #2}

\def\PublisherCambridge{Cambridge: Cambridge University Press}
\def\PublisherKluwer{Dordrecht: Kluwer Academic Publishers}
\def\PublisherASP{San Francisco: ASP}
\def\PublisherReidel{Dordrecht: D. Reidel Publishing Company}

\def\PublisherUAP{Tokyo: Universal Academy Press}

\begin{document}
\SetRunningHead{T. Kato et al.}{V1494 Aql: Eclipsing Fast Nova}

\Received{}
\Accepted{}

\title{V1494 Aql: Eclipsing Fast Nova with an Unusual Orbital Light Curve}

\author{Taichi \textsc{Kato}, Ryoko \textsc{Ishioka}, Makoto \textsc{Uemura}}
\affil{Department of Astronomy, Kyoto University,
       Sakyo-ku, Kyoto 606-8502}
\email{tkato@kusastro.kyoto-u.ac.jp, ishioka@kusastro.kyoto-u.ac.jp, uemura@kusastro.kyoto-u.ac.jp}

\author{Donn R. \textsc{Starkey}}
\affil{AAVSO, 2507 County Road 60, Auburn,, Auburn, Indiana 46706, USA}
\email{starkey73@mchsi.com}

\email{\rm{and}}

\author{Tom \textsc{Krajci}}
\affil{1688 Cross Bow Circle, Clovis, New Mexico 88101, USA}
\email{tkrajci@san.osd.mil}



\maketitle

\begin{abstract}
   We present time-resolved photometry of V1494 Aql (Nova Aql 1999 No. 2)
between 2001 November and 2003 June.  The object is confirmed to be an
eclipsing nova with a period of 0.1346138(2) d.  The eclipses were present
in all observed epochs.  The orbital light curve shows a rather
unusual profile, consisting of a bump-like feature at phase 0.6--0.7
and a dip-like feature at phase 0.2--0.4.  These features were probably
persistently present in all available observations between
2001 and 2003.  A period analysis outside the eclipses has confirmed
that these variations have a period common to the orbital period, and
are unlikely interpreted as superhumps.  We suspect that structure
(probably in the accretion disk) fixed in the binary rotational frame
is somehow responsible for this feature.
\end{abstract}

\section{Introduction}

   Classical novae outbursts are thermonuclear runaways
(TNR; cf., \cite{sta87novareview}; \cite{sta99novareview};
\cite{sta00novareview})
on a mass-accreting white dwarf in cataclysmic variables (CVs)
[for a general review of CVs, see \cite{war95book}].

   Some old novae were later found to be eclipsing.  The classical
examples include
T Aur (Nova Aur 1891, orbital period ($P_{\rm orb}$) = 0.204378 d,
\cite{wal62taur}; \cite{wal63taur}),
DQ Her (Nova Her 1934, $P_{\rm orb}$ = 0.193621 d,
\cite{wal54dqher}; \cite{wal55dqher}; \cite{wal56dqher}),
BT Mon (Nova Mon 1939, $P_{\rm orb}$ = 0.333814 d, \cite{rob82btmon}),
V Per (Nova Per 1887, $P_{\rm orb}$ = 0.10712 d, \cite{sha89vper}),
WY Sge (Nova Sge 1783, $P_{\rm orb}$ = 0.153635 d, \cite{sha83wysge}).
V1668 Cyg (Nova Cyg 1978, $P_{\rm orb}$ = 0.1384 d, \cite{kal90v1668cyg}
QZ Aur (Nova Aur 1964, $P_{\rm orb}$ = 0.357496 d,
\cite{due87novaatlas}; \cite{szk94v838herqzaur}).

   More recently, systematic searches with deep CCD imaging have
succeeded in detecting more eclipsing classical novae:
DO Aql (Nova Aql 1925, $P_{\rm orb}$ = 0.167762 d, \cite{sha93doaqlv849oph}),
RR Cha (Nova Cha 1953, $P_{\rm orb}$ = 0.1401 d,
\cite{wou02rscarv365carv436carapcrurrchabioricmphev522sgr}),
BY Cir (Nova Cir 1995, $P_{\rm orb}$ = 0.282 d, \cite{wou03tvcrv}),
DD Cir (Nova Cir 1999, $P_{\rm orb}$ = 0.0975 d, \cite{wou03tvcrv}),
CP Cru (Nova Cru 1996, $P_{\rm orb}$ = 0.946 d, \cite{wou03tvcrv}),
V849 Oph (Nova Oph 1919, $P_{\rm orb}$ = 0.172755 d, \cite{sha93doaqlv849oph}),
V630 Sgr (Nova Sgr 1936, $P_{\rm orb}$ = 0.1180 d,
\cite{wou01v359cenxzeriyytel}),
QU Vul (Nova Vul 1984 No. 2, $P_{\rm orb}$ = 0.111765 d, \cite{mis95quvul};
\cite{sha95quvul}).

   Although these objects have provided a wealth of knowledge in secular
(\cite{pri75CVperiodchange}; \cite{beu84CVperiodchange})
and nova-induced (\cite{sch83novaPorb})
period changes in classical novae, post-nova accretion disks
(\cite{woo92vpereclipsemapping}; \cite{whi96btmon}), and CVs in general
(\cite{hel00swsexreview}),
the eclipsing nature of these object, however, was discovered long after
their nova outbursts.

   V838 Her (Nova Her 1991, $P_{\rm orb}$ = 0.297635 d) has been the only
object whose eclipsing nature was recognized early during the nova outburst
(\cite{kat91v838heriauc}; \cite{lei93v838her}; \cite{szk94v838herqzaur};
\cite{ing95v838her}).  The detection of eclipses during nova outburst
provides unique opportunity in determining the structure of the outbursting
nova and accretion disk.  This advantage has been recently best
demonstrated in eclipsing recurrent novae:
U Sco (\cite{hac00uscoburst}; \cite{hac00uscoqui}; \cite{mat03usco};
\cite{mun99usco}),
CI Aql (\cite{mat01ciaql}; \cite{hac01ciaql}; \cite{hac03ciaqlmodel}),
and IM Nor (\cite{wou03imnor}).

   V1494 Aql (=Nova Aql 1999 No. 2) is a bright classical nova
($V\sim$4.0 at maximum) discovered by A. Pereira \citep{per99v1494aqliauc}.
\citet{fuj99v1494aqliauc7324} and \citet{mor99v1494aqliauc7325} reported
early optical spectroscopy confirming the nova nature of this object.
\citet{pon99v1494aqliauc7330} reported sub-mm detections with SCUBA.
Further spectroscopy was reported by \citet{kis00v1494aql},
\citet{anu01v1494aql} and \citet{ark02v1494aql}.
\citet{kaw01v1494aql} reported the evidence of an asymmetric outburst
from the presence of significant intrinsic polarization.
\citet{iij03v1494aql} reported further detailed spectroscopy, which
implied the presence of high-velocity jets.
\citet{dra03v1494aqlXray} reported the discovery of X-ray pulsations
with a period of 2523 s, which were ascribed to the first evidence of
$g$-mode pulsations of the white dwarf in an outbursting nova.

   The first suggestion of periodic short-term modulations was made
by \citet{nov00v1494aqliauc}, who reported (from observations 2000
June 7--16) 0.03 mag variations with a period of 0.0627(1) d.
\citet{ret00v1494aqliauc} further suggested, from observations during
2000 June--August, a periodicity of 0.13467(2) d.  The full amplitude
of the variation grew to 0.07 mag in August.  The light curve was
reported to be composed of double-wave modulations.
\citet{bos01v1494aqliauc} reported that the amplitude of the variation
in 2001 June--July grew to 0.5 mag, and suggested that the light
variation was caused by partial eclipse of the accretion disk.
\citet{bar03v1494aql} refined, from observations in 2002 July and
September, the period to be 0.1346141(5) d.

\begin{figure*}
  \begin{center}
    \FigureFile(160mm,100mm){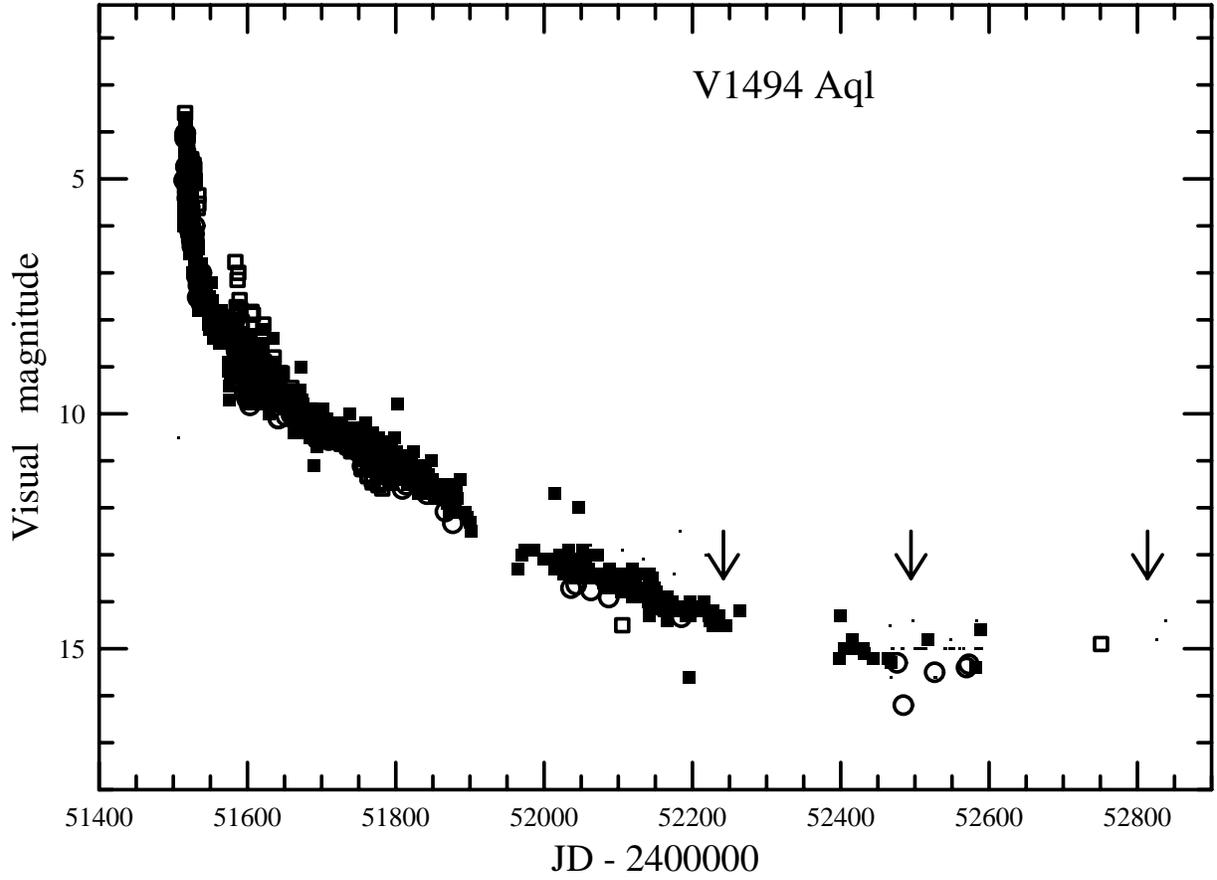}
  \end{center}
  \caption{Light curve of V1494 Aql from observations to VSNET.
  The large filled squares and small dots represent visual magnitudes
  and visual upper limit observations, respectively.  The open circles
  and open squares represent CCD or photoelectric $V$ and $R_{\rm c}$
  observations.  The three arrows represent the mean epochs of our
  time-resolved observations in 2001, 2002 and 2003.}
  \label{fig:lc}
\end{figure*}

\begin{figure}
  \begin{center}
    \FigureFile(88mm,60mm){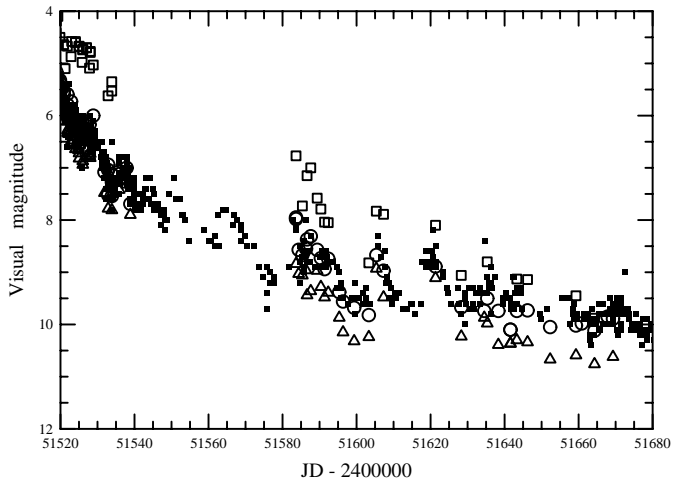}
  \end{center}
  \caption{Enlargement of light curve of V1494 Aql during the oscillation
  stage.  The symbols are the same as in figure \ref{fig:lc},
  supplemented with $B$-band data (open triangles).
  }
  \label{fig:trans}
\end{figure}

   Figure \ref{fig:lc} shows the light curve of V1494 Aql from observations
reported to VSNET \citep{VSNET}.\footnote{
$\langle$http://www.kusastro.kyoto-u.ac.jp/vsnet/$\rangle$.
}  The three arrows represent the mean epochs of our time-resolved
observations (section \ref{sec:ccdobs}) in 2001, 2002 and 2003.
The nova exhibited strong transition-phase oscillations between JD 2451530
(late 1999 December) and 2451650 (2000 April).  Although early stage of
this oscillation phase was presented in \citet{kis00v1494aql}, we
here provides a multicolor light curve covering the entire oscillation
phase in figure \ref{fig:trans}.  This oscillation phase is observed
in a certain fraction of classical novae \citep{GalacticNovae}, for
which an interpretation as the intrinsic instability in a porous
super-Eddington wind has been recently proposed
(\cite{sha01radativehydroinstability};
\cite{sha01novasuperEddington}; \cite{sha02superEddproc}).

\section{CCD Time-Resolved Observation}\label{sec:ccdobs}

   On six nights in 2001 November--December, one night in 2002 August,
and six nights in 2003 June, we obtained CCD time-resolved photometry
of this nova through the VSNET Collaboration.
Compared to the optical maximum, the nova was fainter by 10, 11 and 12 mag,
respectively, at the times of these observations.
The observations were done with unfiltered CCD cameras.  Primary and
secondary local comparison stars were GSC 473.4227 and GSC 473.4367.
The observed magnitudes were first reduced relative to the primary local
comparison star, and were subtracted for nightly averages and slow
long-term variations (less than 0.04 mag d$^{-1}$).  The observed system
was close to $R_{\rm c}$.  The errors of single measurements were typically
less than 0.01--0.03 mag for Starkey and Krajci, and 0.07--0.14 mag for
the Kyoto observations.  The observers' details and log of observations
are given in table \ref{tab:equipment} and \ref{tab:log}, respectively.
Krajci's observations were performed in Tashkent, Uzbekistan.
The Kyoto observations were analyzed a Java$^{\rm TM}$-based PSF
photometry package developed by one of the authors (TK).
Barycentric corrections to the observed times were applied before the
following analysis.

\begin{table}
\caption{Observers and Equipment.}\label{tab:equipment}
\begin{center}
\begin{tabular}{cccc}
\hline\hline
Observer    & Telescope\commenta &  CCD  & Software \\
\hline
Kyoto       & 30-cm SCT & ST-7E   & Java \\
Starkey     & 36-cm SCT & SX-10XE & AIP4Win \\
Krajci      & 28-cm SCT & ST-7E   & AIP4Win \\
\hline
 \multicolumn{4}{l}{\commenta SCT = Schmidt-Cassegrain telescope.} \\
\end{tabular}
\end{center}
\end{table}

\begin{table*}
\caption{Journal of CCD photometry.}\label{tab:log}
\begin{center}
\begin{tabular}{ccrccrc}
\hline\hline
\multicolumn{3}{c}{Date}& Start--End\commenta & Exp(s) & $N$ & Observer \\
\hline
2001 & November & 25 & 52238.919--52238.929 &  30 &  25 & Kyoto \\
     &          & 28 & 52241.855--52241.946 &  30 & 218 & Kyoto \\
     &          & 30 & 52243.860--52243.872 &  30 &  30 & Kyoto \\
     & December &  1 & 52244.867--52244.879 &  30 &  28 & Kyoto \\
     &          &  2 & 52245.890--52245.898 &  30 &  19 & Kyoto \\
     &          &  4 & 52247.866--52247.882 &  30 &  36 & Kyoto \\
2002 & August   &  8 & 52494.651--52494.771 &  90 &  98 & Starkey \\
2003 &  June    & 20 & 52810.654--52810.793 &  90 & 118 & Starkey \\
     &          & 21 & 52811.668--52811.820 &  90 & 134 & Starkey \\
     &          & 22 & 52812.656--52812.880 &  90 & 186 & Starkey \\
     &          & 22 & 52813.322--52813.457 & 240 &  48 & Krajci \\
     &          & 23 & 52813.669--52813.882 &  90 & 182 & Starkey \\
     &          & 23 & 52814.262--52814.456 & 240 &  69 & Krajci \\
     &          & 24 & 52815.249--52815.458 & 240 &  74 & Krajci \\
     &          & 25 & 52815.684--52815.888 &  90 & 174 & Starkey \\
\hline
 \multicolumn{7}{l}{\commenta BJD$-$2400000.} \\
\end{tabular}
\end{center}
\end{table*}

\section{Eclipses}

   We determined mid-eclipse times by minimizing the dispersions of eclipse
light curves folded at the mid-eclipse times.  The error of eclipse times
were estimated using the Lafler--Kinman class of methods, as applied by
\citet{fer89error}.  As shown later, the mean eclipse light curve has
a small degree of asymmetry.  The error estimates should therefore be
treated as a statistical measure of the observational errors (see also
\cite{kat02ircom} for more discussion of this application to CV eclipses).

   The determined times of minima are given in table \ref{tab:eclmin},
as well as the minimum times from \citet{bar03v1494aql}.  The initial
eclipse in \citet{bar03v1494aql} is defined as $E$ = 0.
The BJD and HJD agree within 0.0001 d at the observed epochs of
\citet{bar03v1494aql}.  Although individual error estimates were not
listed in \citet{bar03v1494aql}, an overall uncertainty of 0.0002 d was
given.  A linear regression of the observed eclipse times (the 2001
eclipse was excluded from this regression because of the incomplete
coverage of the eclipse and the rather large $O-C$ against the rest of
the observations) yielded the following ephemeris (the quoted
errors represent 1$\sigma$ errors at $E$ = 1467):

\begin{equation}
\rm{BJD_{min}} = 2452458.3230(3) + 0.1346138(2) $E$. \label{equ:reg1}
\end{equation}

   We thus obtained a refined period of 0.1346138(2) d.  We consistently
use this ephemeris throughout the following discussion.

\begin{table}
\caption{Eclipse minima.}\label{tab:eclmin}
\begin{center}
\begin{tabular}{ccccc}
\hline\hline
BJD\commenta & Error\commentb & $E$ & $O-C$\commentb\commentc 
             & Source\commentd \\
\hline
52241.8604 &        4 & $-$1608 & $-$36 & 1 \\
52458.324  & $\cdots$ &    0 &    10 & 2 \\
52458.4580 & $\cdots$ &    1 &     4 & 2 \\
52462.4967 & $\cdots$ &   31 &     7 & 2 \\
52464.5164 & $\cdots$ &   46 &    12 & 2 \\
52471.513  & $\cdots$ &   98 & $-$21 & 2 \\
52473.533  & $\cdots$ &  113 & $-$13 & 2 \\
52494.6670 &       10 &  270 & $-$17 & 1 \\
52519.3042 & $\cdots$ &  453 &    12 & 2 \\
52520.2460 & $\cdots$ &  460 &     7 & 2 \\
52810.7410 &        3 & 2618 &  $-$9 & 1 \\
52811.6840 &        2 & 2625 &  $-$2 & 1 \\
52812.7617 &        2 & 2633 &     5 & 1 \\
52813.4344 &       10 & 2638 &     2 & 1 \\
52813.7050 &        2 & 2640 &    16 & 1 \\
52813.8392 &        4 & 2641 &    11 & 1 \\
52814.3756 &        4 & 2645 &  $-$9 & 1 \\
52815.3184 &        5 & 2652 &  $-$4 & 1 \\
52815.7217 &        4 & 2655 & $-$10 & 1 \\
52815.8573 &        2 & 2656 &     0 & 1 \\
\hline
 \multicolumn{5}{l}{\commenta BJD$-$2400000.} \\
 \multicolumn{5}{l}{\commentb Unit 0.0001 d.} \\
 \multicolumn{5}{l}{\commentc Against equation (\ref{equ:reg1}).} \\
 \multicolumn{5}{l}{\commentd 1: this work, 2: \citet{bar03v1494aql}.} \\
\end{tabular}
\end{center}
\end{table}

\section{Orbital Light Curve}

\begin{figure}
  \begin{center}
    \FigureFile(88mm,120mm){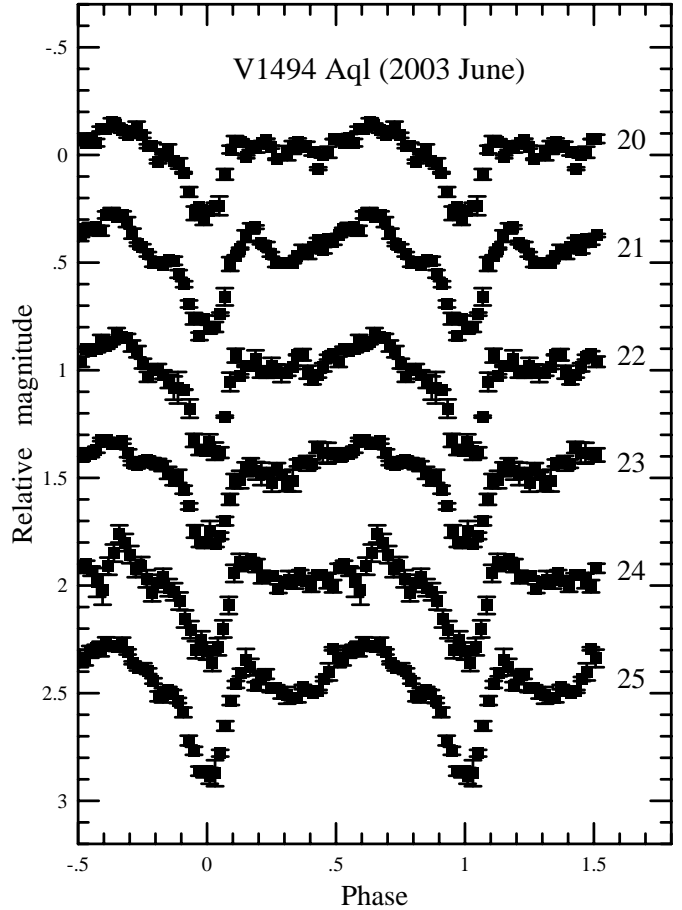}
  \end{center}
  \caption{Nightly averaged orbital light curves in 2003 June.
  The orbital phase was calculated against equation (\ref{equ:reg1}).
  The numbers in the right side represent the calendar day of June.
  }
  \label{fig:ph}
\end{figure}

   Figure \ref{fig:ph} presents nightly averaged orbital light curves in
2003 June.  Outside eclipse around phase = 0, there was a bump-like feature
at phase 0.6--0.7, and a dip-like feature at phase 0.2--0.4 on almost
all nights.  The relative stability of the phases of these features
indicate that they are unlikely features associated with superhumps
(\cite{vog80suumastars}; \cite{war85suuma}; \cite{pat99SH};
see also \cite{ret97v1974cygSH}; \cite{ski97v1974cyg}; \cite{ole02v1974cyg}
for an example and discussion of well-observed superhumps in a fading
classical nova V1974 Cyg).  The same features, although the amplitudes
were smaller, can be also seen in the light curves in \citet{bar03v1494aql},
suggesting that these features have been persistently present.

\begin{figure}
  \begin{center}
    \FigureFile(88mm,120mm){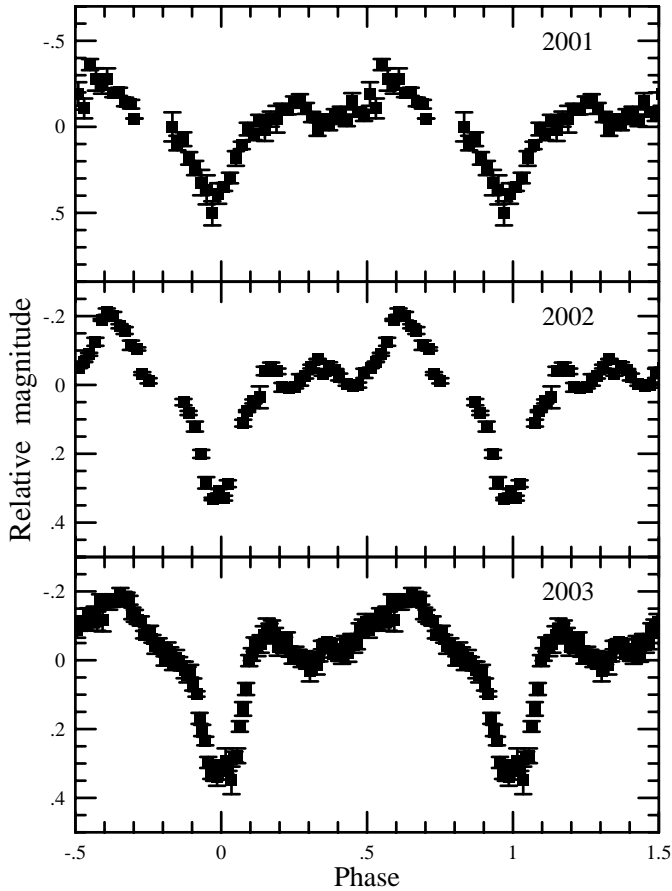}
  \end{center}
  \caption{Comparison of the mean orbital light curves between 2001, 2002
  and 2003 (note that the 2002 light curve was constructed from a single-night
  observation on 2002 August 8, and that the orbital phase in the 2001
  light curve was an extrapolated values; the magnitude scale is also
  different for the 2001 light curve).  The overall appearance of
  out-of-eclipse features are common between these three epochs.
  }
  \label{fig:ave}
\end{figure}

   Figure \ref{fig:ave} presents a comparison of the mean orbital light
curves between 2001, 2002 and 2003 [note that the mean light curve
for 2001 is based on limited phase coverage (see table \ref{tab:log}),
and the mean  light curve for 2002 is in fact based on a single night of
data].  The overall appearance of out-of-eclipse
features are common between these two epochs, confirming the tendency
seen in the 2002 light curves \citep{bar03v1494aql}.  These features were
already present in 2001.

\begin{figure}
  \begin{center}
    \FigureFile(88mm,60mm){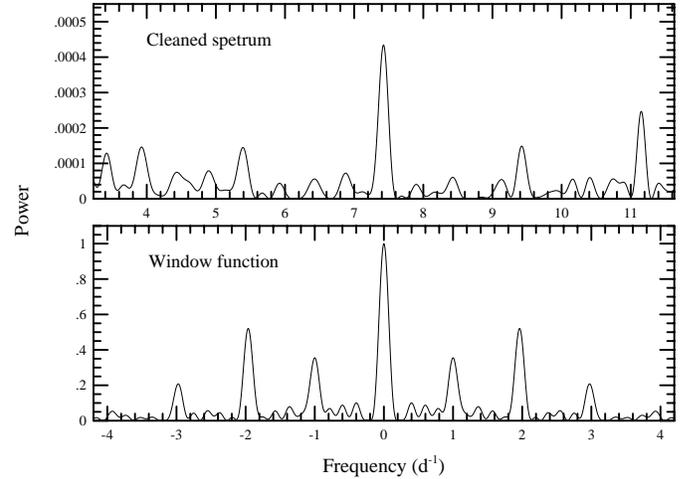}
  \end{center}
  \caption{(Upper:) Cleaned power spectrum outside the eclipses.  The powers
  are given in an arbitrary unit.  The only significant periodicity is
  at a frequency of 7.43 d$^{-1}$, which is identical with the orbital
  period.  The peak near frequencies 5.4 and 9.4 d$^{-1}$ are side-lobes
  arising from the observation windows.
  (Lower:) Window function.  These diagrams are from the 2003 data.
  }
  \label{fig:clean}
\end{figure}

   Figure \ref{fig:clean} shows period analysis of V1494 Aql with the Clean
method \citep{CLEAN}, with a gain parameter of 0.01.  The data were limited
to $|phase|\geq$0.2 to avoid the effect from the eclipses.  The only
significant periodicity is at a frequency of 7.43 d$^{-1}$, which is
identical with the orbital period.  The peak near frequencies 5.4 and 9.4
d$^{-1}$ are side-lobes arising from the observation window.  The reality
of a possible weak signal around frequency 11.2 d$^{-1}$ was not confirmed
by analysis of subdivided data into 2--3 d lengths.
Within 40\% of the orbital period (which is a safe limit
for the known superhumps: \cite{pat99SH}), there is no indication of
superhumps.  The result also seems to preclude the intermediate polar-type
(cf. \cite{kin90IP}; \cite{pat94ipreview}; \cite{hel96IPreview})
magnetically controlled accretion, which was proposed to be the cause
of oscillations in transition-phase novae (\cite{ret02novaoscillationproc}).

   The double-wave orbital light curves bear resemblance to those
of luminous supersoft X-ray sources (SSXS: \cite{kah97SSSreview});
this phenomenological association may look reasonable in the light
of supersoft X-ray detection in V1494 Aql during its post-outburst
state \citet{dra03v1494aqlXray}.
The light curves of SSXS almost always show, however, brighter maximum
at phase 0.2--0.4 than phase 0.6--0.8:
QR And: \citet{beu95qrand}; \cite{mat96qrand}; \cite{mey98qrand}),
V Sge: \citet{sim96vsge}; \citet{sim99vsge}; \citet{ste98vsgestars};
\citet{pat98vsgetpyx}; \citet{sim00vsge},
CAL 87: \citet{cal89cal87}; \citet{alc97cal87}; \citet{hut98cal87}.
Model calculations (\cite{sch97cal87}; \cite{mey97SSSdiskrim})
suggest that this phenomenon can be reproduced by considering the
thickening of the accretion disk rim near the stream impact point.
This explanation is less likely applicable to the features observed
in V1494 Aql.

   Some eclipsing polars occasionally show similar light curves with
similar bumps and dips: e.g. V2301 Oph (\cite{bar94v2301oph}),
MN Hya (\cite{sek94mnhya}), HU Aqr (\cite{sch93huaqr}).  In these polars,
magnetically controlled accretion on a magnetic white dwarf synchronously
rotating with orbital motion of the secondary.  This possibility might
be attractive because of the phase stability of the observed features.
However, the quiescent magnitude of V1494 Aql more suggests a normal
CV with an accretion disk (\cite{kis00v1494aql}; \cite{war86NLabsmag}),
which would almost preclude the polar-type interpretation.

   Among the known class of luminous (i.e. with a hot accretion disk)
CVs, double-wave orbital modulations are known to be present during
the early stage of superoutbursts of WZ Sge-type dwarf novae
(\cite{bai79wzsge}; \cite{dow81wzsge}; \cite{kat01hvvir}).  These
modulations are called either {\it early superhumps}
(\cite{kat01hvvir}; \cite{kat02wzsgeESH}; \cite{ish02wzsgeletter}),
{\it early humps} (\cite{osa02wzsgehump}; \cite{osa03DNoutburst}) or
{\it outburst orbital humps} (\cite{pat02wzsge}), although the last
nomenclature is apparently mislabeled in that it was from the
misinterpretation of the observation as an enhanced hot spot,
which was originally claimed to explain the 1978 outburst of WZ Sge
\citep{pat81wzsge} (see \cite{osa03DNoutburst} for a discussion).

   From the stability of the orbital (out-of-eclipse) light curve over
years, we suspect that structure (probably in the accretion disk)
fixed in the binary rotational frame, as in early superhumps of
WZ Sge-type dwarf novae, is somehow responsible for the observed
light curve.

\section{Summary}

   We present long-term and time-resolved photometry of V1494 Aql
(Nova Aql 1999 No. 2) based on VSNET observations.  The time-resolved
photometry, undertaken in 2001 November--December, 2002 August and
2003 June, confirmed that the object is an eclipsing nova with
a period of 0.1346138(2) d.
The object is a rare classical nova whose eclipsing nature was
recognized during the decline stage of a nova outburst.
The eclipses were equally present in all 2001, 2002 and 2003
observations.  The orbital light curve shows a rather unusual profile,
consisting of a bump-like feature at phase 0.6--0.7 and a dip-like
feature at phase 0.2--0.4.  These features were present in almost
all our observations and in those in the literature between
2001 and 2003.
A period analysis outside the eclipses has confirmed that these
variations have a period common to the orbital period, and are
unlikely interpreted as superhumps.  The double-wave modulation
somewhat resembles those of supersoft X-ray sources, but the profile
in V1494 Aql is different from those of supersoft X-ray source
in its primary maximum occurring at phase 0.6--0.7.
We suspect that structure (probably in the accretion disk)
fixed in the binary rotational frame is somehow responsible for
this feature.

\vskip 3mm

We are grateful to many observers who have reported vital observations
to VSNET.  We are grateful to Izumi Hachisu for helpful discussion.
This work is partly supported by a grant-in-aid (13640239, 15037205)
from the Ministry of Education, Culture, Sports, Science and Technology.

\end{document}